# Human-Assisted Graph Search: It's Okay to Ask Questions

Aditya Parameswaran
Stanford University
adityagp@cs.stanford.edu

Anish Das Sarma
Yahoo! Research
anishdas@yahoo-inc.com

Hector Garcia-Molina
Stanford University
hector@cs.stanford.edu

Neoklis Polyzotis
UC Santa Cruz
alkis@cs.ucsc.edu

Jennifer Widom
Stanford University
widom@cs.stanford.edu

## ABSTRACT

We consider the problem of *human-assisted graph search*: given a directed acyclic graph with some (unknown) target node(s), we consider the problem of finding the target node(s) by asking an omniscient human questions of the form "Is there a target node that is reachable from the current node?". This general problem has applications in many domains that can utilize human intelligence, including curation of hierarchies, debugging workflows, image segmentation and categorization, interactive search and filter synthesis. To our knowledge, this work provides the first formal algorithmic study of the optimization of human computation for this problem. We study various dimensions of the problem space, providing algorithms and complexity results. We also compare the performance of our algorithm against other algorithms, for the problem of webpage categorization on a real taxonomy. Our framework and algorithms can be used in the design of an optimizer for crowd-sourcing platforms such as Mechanical Turk.

## 1. INTRODUCTION

*Crowd-sourcing* services, such as Amazon's Mechanical Turk (mturk.com) and CrowdFlower (crowdflower.com) allow organizations to set up tasks that humans can perform for a certain reward. The goal is to harness "human computation" in order to solve problems that are very difficult to tackle completely algorithmically. Examples of such problems in practice include object recognition, language understanding, text summarization, ranking, and labeling [9].

In a typical crowd-sourcing setting, the tasks are broken down to simple questions (often with a YES/NO answer) that can be easily tackled by humans. Since each question comes at a price, be it money, effort, or time, it is desirable to minimize the number of questions that need to be answered in order to achieve the overall objective. Thus, we would like a general-purpose *human computation optimizer* that selects the specific questions to be asked so as to minimize some cost metric. (The vision for such an optimizer, leveraging human and algorithmic computation along with relational data was laid out recently [3].) We develop core algorithms for such an optimizer, considering a class of *human-assisted graph search* problems.

In human-assisted graph search, or HumanGS for short, we are given as input a directed acyclic graph that contains a set of unknown *target* nodes (collectively called the *target set*), and the goal is to discover the identities of these target nodes solely by asking *search questions* to humans. A search question is a question of the form "Given a node $x$ in the graph, is there a target node reachable from node $x$ via a directed path?". (Note that, by definition, each search question corresponds to a node in the graph, and we can ask a search question at any node.) The objective is to select the optimal set of nodes at which to ask search questions in order to ascertain the identities of the target nodes. Later, we examine different notions of optimality, including minimizing the total number of questions, or minimizing the set of resulting possible target nodes after a fixed number of questions.

EXAMPLE 1.1. *Suppose our goal is to categorize an image into one of the classes of the hierarchical taxonomy shown in Figure 1(a). If the image is that of a Nissan car, but the model is not identifiable, the most suitable category is "**Nissan**". If the model is identifiable as well, say, Sentra, then the most suitable category would be "**Sentra**". Since categorization of images is a task that can be performed better by humans than by computers, we would like to utilize human intelligence. We wish to ascertain the most suitable category (which could be anywhere in the taxonomy) by asking the minimum number of questions to humans.*

*This task is an instance of the HumanGS problem. The taxonomy is the DAG, with each category corresponding to a node. The target node is the most suitable category of the image. Asking a search question at a node/category **X** is equivalent to asking a question of the form "Is this an **X**?". For instance, in the Nissan car example above, receiving a YES answer to "Is this a **Car**?" says that the most suitable category, in this case **Nissan**, is reachable from the category **Car**. Also, asking a question at the root, **Vehicle**, (a general question) gives a YES answer while asking a question at a leaf, **Maxima** (a specific question), gives a NO answer. If the image is that of a car, but the model is not identifiable, then asking a question at **Vehicle** and **Car** will yield YES, while all other questions will receive a NO answer.* □

There are several interesting properties that make HumanGS a nontrivial problem. First, the answers to different search questions may be correlated, e.g., if the answer to the search question at a node is YES, then the answer to a search question at an ancestor of that node will be YES as well. Therefore, it is possible to identify the target nodes without asking search questions at all nodes in the graph. Second, the location of a node affects the amount of information that can be obtained from the corresponding search question. Asking search questions at nodes close to leaves (very *specific* questions) are more likely to receive a negative answer, while asking questions at nodes close to roots (very *general* questions) are more likely to receive positive answers. Asking search questions at the "middle" nodes may give more information. In this sense, the HumanGS problem is similar to the 20 questions game[1], where very specific or very general questions do not help.

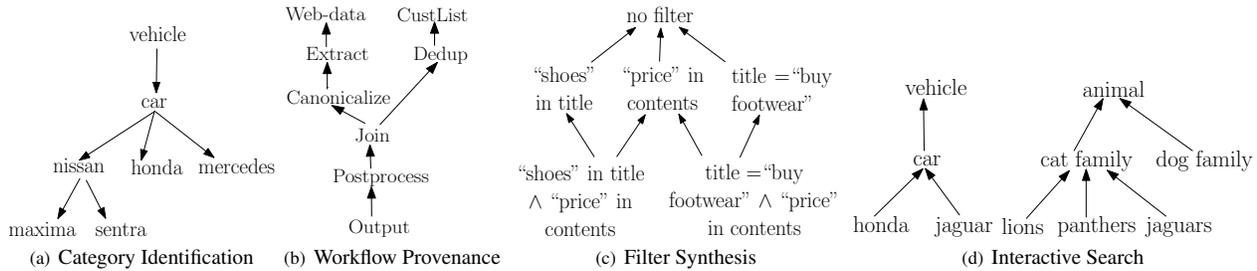

**Figure 1: Examples of HumanGS**

An additional challenge stems from the usage model of crowd-sourcing services in practice. Ideally, we would like to issue one question at a time, selecting the next search question based on the answers to previous questions. However, typical crowd-sourcing services incur a high latency for obtaining the answer to a single question: the question first has to be posted to the service, then a human worker has to find the question and decide to answer it, and finally the answer is sent back to the requester. This characteristic leads one away from the sequential one-question-at-a-time model. A more common model is to issue several questions in parallel, whose answers are then combined to solve the task at hand. The challenge, therefore, is to reason about the possible answers for search questions at different nodes in the graph, and to select the set of questions in order to infer as much information as possible about the target nodes *across all possibilities*. In this paper we study this "offline" model analytically. However, in Section 6 (Experiments) we briefly look at a hybrid approach, where we ask some questions, examine the results, and then ask additional questions, but a full theoretical study of a hybrid model is left for future work.

In this paper, we develop algorithms that compute the optimal set of questions for different variants of the HumanGS problem. Our primary contributions can be summarized as follows:

- We identify a number of real-world instances of the HumanGS problem, which lead to the delineation of three orthogonal problem dimensions. (Section 2)
- We formally define the HumanGS problem. (Section 3)
- We present algorithms and complexity results for the problem for the three dimensions identified in Section 2. We show that while the general problem is computationally hard, the more constrained variants are tractable. (Sections 4 and 5)
- We study the performance of our algorithms versus others for an instance of HumanGS on real-world data. (Section 6)

To the best of our knowledge, ours is one of the first papers to address the problem of optimizing a computation that harnesses human "processors" through a crowd-sourcing service. Requesting input from humans as a component of a computer algorithm is not new; the field of active learning [6] also considers the problem of optimally requesting input from experts. However, this input is typically for generating test datasets for machine learning problems.

Other studies examine different aspects of human computation technologies, such as social issues and application usage. HumanGS bears some resemblance to classification tasks solved by decision trees, but in our case no training set, statistics, or attributes are provided, and the only questions that can be asked involve reachability. We discuss related work in more detail in Section 7.

---

[1]20 questions is a two player game, where one player thinks of an object, person or place and the other player has to guess the identity of that item by asking the first player up to 20 YES/NO questions.

## 2. APPLICATIONS AND DIMENSIONS

We will study variants of the HumanGS problem along three orthogonal dimensions. The choice of these dimensions is driven by applications that represent instantiations of HumanGS in practice. We discuss the applications first and then formally define the derived dimensions.

- **Image Categorization**: Described in Example 1.1.
- **Manual Curation**: (Isomorphic to image categorization.) In an existing taxonomy (such as Wikipedia, phylogenetic trees, web of concepts [13]), we wish to manually insert new concepts, topics and items to their most suitable location in the hierarchy. Manual curation of each new item can be reduced to an instance of HumanGS in a manner similar to image categorization. A search question at $x$ corresponds to the question "Is the item a kind of $x$?", while the target node is the most suitable parent of the new item.
- **Debugging of Workflows**: Suppose that we detect an erroneous result in the output of a workflow. Naturally, we would like to detect the earliest workflow steps that introduced the error. Assume that we maintain provenance information for the workflow [17, 21], so that we can identify the fragment of the output of each workflow step that is linked to the erroneous result. If we view the workflow as a DAG (with the direction of the edges reversed), as in Figure 1(b), then isolating the earliest incorrect workflow steps can be reduced to an instance of HumanGS. A search question at $x$ corresponds to asking the user "Is the output fragment at point $x$ wrong?", and the target nodes are the earliest steps in the workflow that introduced errors into the resulting output.
- **Filter Synthesis**: Suppose that a user wishes to apply a filter on a data set as part of some ad-hoc analysis. We can help the user formulate the filter by asking them a few questions. If we arrange the candidate filters in a DAG as in Figure 1(c), (Candidate filters can be extracted by analyzing the data set [1]) then filter synthesis can be reduced to an instance of HumanGS. Edges in the DAG indicate logical implication between filter fragments. A search question at $x$ corresponds to asking the user "Do you want all data items satisfying condition $x$ to be part of the result?". The target set comprises all filters whose disjunction yields the intended filter.
- **Interactive Search**: In interactive search [8, 15], the search engine asks the user a few questions that help isolate the concepts that best encompass his/her information need. The questions are based on a backend hierarchy of concepts that cover the crawled web-data. Figure 1(d) shows an example hierarchy. (Note that the edges in the hierarchy go from more specific to more general concepts.) After presenting the user with initial results, the search engine can pose questions of the form "Do you want more results like concept X?". The target nodes are the most general concepts that the user is interested in (i.e., those that encompass his information need).

Inspired by these applications, we derive three dimensions that characterize the different instances of the HumanGS problem.

**Dimension 1: Single/Multi**.

The first dimension controls the characteristics of the target set. In the Single variant, the target set contains a single node. The

268

Multi variant does not constrain the size of the target set.

While Single is relevant for Image Categorization and Manual Curation, the Multi variant is relevant for the other three applications listed above. As a concrete example, for a given incorrect result at the output in Figure 1(b), both the canonicalization step on web-data as well as the deduping of the customer list could be introducing errors. As another example, consider creating a filter for web-pages relating to shoe shopping, as in Figure 1(c). Here, a user might be interested in pages that contain "shoes" in the title and "price" in the contents, or pages that have as a title "buy footwear".

**Dimension 2: Bounded/Unlimited**.

The second dimension controls the number of questions that can be asked. In the Bounded case, we are given a budget $k$ for the total number of questions that can be asked and we want to compute a node set $N$, $|N| \leq k$, at which to ask search questions such that we narrow down the candidates for the target set as much as possible. The Unlimited case does not put a bound on the number of questions. In this case, we want to compute the minimal set of nodes to ask questions such that we precisely identify the target set.

The Bounded case is relevant when asking questions is potentially costly, e.g., on Mechanical Turk, and we wish to bound the total cost while narrowing down the possibilities for the target nodes. The Bounded case is also relevant in Interactive Search, where it is not practical to ask an unlimited number of questions (since the user may not be willing to answer too many questions). Once we receive the answers to the questions, we can display results related to the concepts that may correspond to target nodes. The results can still be useful for the user, even if we do not identify the target set precisely. The Unlimited case is relevant whenever an exact answer is required, such as Manual Curation of hierarchies.

**Dimension 3: DAG/Downward-Forest/Upward-Forest.**

The third dimension controls the type of DAG on which we perform the search. Besides a general DAG, which is relevant for Filter Synthesis (as in Figure 1(c)) and Debugging of Workflows, we consider two restricted structures. The first is a "downward forest" structure, where there are several trees with edges directed from parents to children, as in Figure 1(a). This structure is relevant in Image Categorization and Manual Curation. The second structure is an "upward forest", which is the same as a downward-forest except that the edges are reversed, as in Figure 1(d). This structure is relevant in Interactive Search.

## 3. THE HumanGS PROBLEM

Informally, our approach can be summarized by Figure 2 (which illustrates HumanGS for the Single case of Dimension 1.) We first select a set of questions to ask via HumanGS procedure $C$. Subsequently, via an evaluation procedure $E$, we obtain answers from humans to the selected questions, using which we compute the possible candidates for the target node, which can be either the target node or a superset. We now attempt to formalize these intuitions.

We are given a directed acyclic graph $G = (V, E)$ that reflects the semantics of a specific instance of HumanGS. We use $n \equiv |V|$ to denote the number of nodes in the graph. A node $v \in V$ is *reachable* from another node $u \in V$ if there exists a directed path from $u$ to $v$. The *reachable set* of $u$, denoted rset$(u)$, contains all nodes that are reachable from $u$, including $u$. For instance, the reachable set of nissan in Figure 1(a) is {nissan, maxima, sentra}. The *preceding set* of $u$, denoted pset$(u)$, contains all $v \neq u$ such that $u \in$ rset$(v)$. For instance, the preceding set of nissan in Figure 1(a) is {vehicle, car}. We say that $u$ and $v$ are *unrelated* if there is no directed path between them, i.e., $u \notin$ pset$(v) \cup$ rset$(v)$.

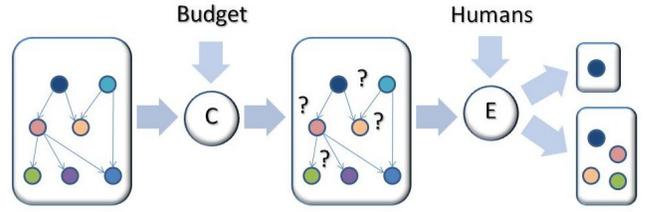

**Figure 2: The HumanGS procedure C selects a set of nodes to ask questions. Once Humans answer the questions via the evaluator procedure E, we may be able to isolate the target node, or a set of nodes one of which is the target node, depending on the answers.**

We assume that the HumanGS instance involves a node-set $U^* \subseteq V$, termed the *target set*, that comprises the *target nodes*. The target set must satisfy the following property:

*Independence Property*: No two nodes in $U^*$ are related.

This property holds in all motivating applications in Section 2. Intuitively, if there are two nodes $u$ and $v \in U^*$ such that $u \neq v$ and $u \in$ rset$(v)$, then $v$ can be discarded because $u$ "subsumes" $v$. For instance, in Figure 1(d), if the user is interested in 'cat family' as well as 'lions', then we would prefer to retain 'cat family' in $U^*$ instead of 'lions' because 'cat family' subsumes 'lions'. (Note that the independence property holds even in the workflow debugging application: if there are multiple related error-causing steps, we identify the earliest ones.)

We can informally describe the HumanGS problem as computing a set of nodes $\{u_1, \ldots, u_k\}$ such that the answers to the corresponding *search questions* at the set of nodes lead to the identification of $U^*$. Note that each search question corresponds to a node in the graph, and hence we interchangeably use "asking a question", "asking a question at a node" and "asking a node". Asking a question is defined formally as follows.

DEFINITION 3.1 (ASKING A QUESTION q$(u, U^*)$). *Asking a question at node $u \in V$, denoted as* q$(u, U^*)$, *returns YES if* rset$(u) \cap U^* \neq \emptyset$, *and NO otherwise.*

In other words, q$(u, U^*)$ returns YES iff a directed path starting at $u$ reaches at least one node in $U^*$. Note that $u$ may itself be in the target set. Also note that for any $U_1^*$ and $U_2^*$, $U_1^* \neq U_2^*$, both of which satisfy the independence property, there is some node at which asking a question would give different answers. (Consider $u$ such that $u \in U_1^*$ and $u \notin U_2^*$. Either there is no $v \in$ rset$(u)$ present in $U_2^*$, in which case asking a question at $u$ would give different answers. Or, there is such a $v \neq u$, in which case asking a question at $v$ would give different answers.)

A solution to HumanGS is always feasible, as we can identify $U^*$ by asking questions at every node in $V$. However, it is not necessary to ask questions at every node, as the following trivial lemma illustrates.

LEMMA 3.2 (DAG PROPERTY). *If* q$(u, U^*)$ *is YES, then* q$(v, U^*)$ *is YES for every $v$ in pset$(u)$. Conversely, if* q$(u, U^*)$ *is NO then* q$(v, U^*)$ *is NO for every node $v$ in rset$(u)$.*

Let $N \subseteq V$ be some set of nodes at which we ask questions. In general, the answers to these questions may not be sufficient to precisely identify $U^*$, since there may be other nodes (not in $U^*$) for which the current answers would be the same even if those nodes *were* in $U^*$. We introduce the notion of a *candidate set* to capture the possibilities for $U^*$ based on questions on a node-set $N$. The candidate set, denoted as cand$(N, U^*)$, is the maximal set of nodes that we cannot distinguish from $U^*$ based solely on the answers to questions at the nodes in $N$. For $|N| = 0$, we have the trivial result that cand$(N, U^*) = V$. We first consider how asking

269

|  |  | **DAG** | **Downward-Forest** | **Upward-Forest** | **Downward Bal.** | **Upward Bal.** |
|---|---|---|---|---|---|---|
| Single | Bounded | NP-Complete$(n,k)$, $O((2n)^k n^2 k)$ | $O(n \log n)$ | $O(mk^2 n^5)$ | $O(1)$ | $O(1)$ |
|  | Unlimited | $\min\{\log^2 o, \log n\} \times$ Single-Bounded | $O(1)$ | $O(n)$ | $O(1)$ | $O(n)$ |
| Multi | Bounded | NP-Hard$(n,k)$, $\Sigma_2^P(n,k)$ | $O(mk^2 n^6)$ | | $O(1)$ | |
|  | Unlimited | | $O(1)$ | | | |

**Table 1: Summary of Results. (Bal. stands for Balanced Trees)**: $k$ is the budget of questions, $n = |V|$, $m$ is the arity of the tree or forest, and $o$ is the size of the optimal $N$.

a single question (i.e., $|N| = 1$) allows us to restrict the contents of the candidate set beyond $V$.

THEOREM 3.3 (ONE QUESTION PRUNING). *Assume that we ask a single question at node $u$. The candidate set is computed as follows, based on the answer and the variant of Single/Multi that the HumanGS instance falls under.*

$$\mathsf{cand}(\{u\}, U^*) = \begin{cases} V - \mathsf{rset}(u) & \mathsf{q}(\{u\}, U^*) = \text{NO} \\ V - \mathsf{pset}(u) & \mathsf{q}(\{u\}, U^*) = \text{YES} \wedge \text{Multi} \\ \mathsf{rset}(u) & \mathsf{q}(\{u\}, U^*) = \text{YES} \wedge \text{Single} \end{cases}$$

The proof can be found in [4]. Given this base case of one question, we can compute $\mathsf{cand}(N, U^*)$ for a general node-set $N$ as the intersection of the candidate sets resulting from individual questions.

THEOREM 3.4. *After asking questions at all nodes in a node-set $N$, we have: $\mathsf{cand}(N, U^*) = \bigcap_{u \in N} \mathsf{cand}(\{u\}, U^*)$.*

The proof can be found in [4]. Thus, each question may enable some additional pruning of the candidate set, and the order in which questions are asked does not affect the final result. As an example, let us consider again the HumanGS problem illustrated in Figure 1(a). Suppose that the single target node is *maxima* (recall that $|U^*| = 1$ for this search task), and assume that we ask questions at $N = \{car, nissan, mercedes\}$. Clearly, the questions at *car* and *nissan* yield YES, whereas the question at *mercedes* yields NO. Based on these answers, we can assert that $\mathsf{cand}(N, U^*) = \{nissan, maxima, sentra\}$. The candidate set contains the target node as well as two "false positives" (*nissan* and *sentra*). Picking $N$ so as to minimize the number of false positives is the goal of the algorithms that we present later.

Since we are operating in an "offline" setting where the answers to previous questions are not provided to us, we are interested in minimizing the size of the candidate set in the worst case. Given that $U^*$ is unknown, we may use the maximum size of $\mathsf{cand}(N, U^*)$ (under all admissible possibilities for $U^*$) as an indication of the worst-case uncertainty that remains after asking the questions in $N$. We use $\mathsf{wcase}(N)$ to denote this worst-case size. A natural objective is to select $N$ so that $\mathsf{wcase}(N)$ is minimized. We define wcase formally in Section 4 for Single and in Section 5 for Multi.

### 3.1 Summary of Results and Outline

Table 1 summarizes our results on the complexity of the examined variants. The details of the analysis and the corresponding algorithms are given in the following sections. The presentation is organized in two sections based on Dimension 1: Single is covered in Section 4, and Multi is covered in Section 5. Each row in the table corresponds to a subsection in the corresponding section. In each case, we first provide a formal definition of the corresponding HumanGS problem, followed by the complexity analysis for the different graph structures. In addition to the structures listed in Dimension 3, we also consider the special case of balanced trees in Appendices A.8, A.11, C.4 and C.5, which admit very efficient solutions.

We omit some proofs due to space constraints. Some of the omitted proofs can be found in the Appendix, while all the proofs can be found in the extended technical report [4].

## 4. SINGLE TARGET NODE

In the Single problem, we have the constraint that there is a single target node, i.e., $|U^*| = 1$. Let this node be $u^*$. To simplify notation for Single, we use $\mathsf{cand}(N, u^*)$, instead of $\mathsf{cand}(N, \{u^*\})$, to denote the candidate set after questions have been asked at the node set $N$, and we use $\mathsf{q}(u, u^*)$, instead of $\mathsf{q}(u, \{u^*\})$, to denote the answer to asking a question at $u$. Recall that as in Theorem 3.3, asking a question at $u$ for the Single problem tells us whether the candidate set $\mathsf{cand}(\{u\}, u^*)$ is $\mathsf{rset}(u)$ (if the answer is YES) or $V - \mathsf{rset}(u)$ (if the answer is NO).

Given a node set $N$ at which we ask questions, we define the *worst-case candidate set size* as

$$\mathsf{wcase}(N) = \max_{u_i \in V} |\mathsf{cand}(N, u_i)| \qquad (1)$$

In other words, wcase computes the size of the largest candidate set when the target node could be any node in $V$.

### 4.1 Single-Bounded

In the Single-Bounded variant we have a fixed budget $k$ on the number of questions that may be asked, i.e., the size of $N$ cannot exceed $k$. The goal is to pick the set of nodes $N$ such that the worst-case candidate set size is minimized.

DEFINITION 4.1 (SINGLE-BOUNDED). *(Bounded Search for a Single target node.) Given a parameter $k$ and the restriction that $|U^*| = 1$, find a set $N$ of nodes $N \subseteq V$ to ask questions such that $|N| = k$ and $\mathsf{wcase}(N)$ is minimized.*

The following subsections examine the complexity of the problem under the different possibilities for the structure of $G$, i.e., general DAG, downward-forest and upward-forest.

#### 4.1.1 Single-Bounded: DAG

We begin with the auxiliary result that $\mathsf{wcase}(N)$ can be computed in time polynomial in the number of nodes in the graph. This result is used later to bound the complexity of Single-Bounded.

THEOREM 4.2 (COMPUTATION OF WORST CASE). *Given a node set $N$ at which questions are asked, $\mathsf{wcase}(N)$ can be computed in $O(n^2 \cdot k)$, where $n = |V|$.*

A formal proof can be found in Appendix A.1. The main idea is to first compute all pairs $(a, b)$ in $V \times V$ such that $b \in \mathsf{rset}(a)$ and then use this information to compute $\mathsf{cand}(N, u^*)$ for every possibility of $u^*$. Using the result above, we can define a brute-force approach to solving Single-Bounded, by considering all possible combinations of $N$ with size at most $k$.

LEMMA 4.3 (BRUTE-FORCE SOLUTION). *The optimal solution of Single-Bounded for any DAG can be found in $O(n^k \cdot n^2 \cdot k)$, where $n$ is the number of nodes in $V$.*

Clearly, we can solve Single-Bounded optimally in PTIME if $k$ is bounded by a constant. However, the appearance of $k$ in the exponent hints at the hardness of the problem in the general case. Indeed, the following result shows that Single-Bounded is computationally hard. The proof of the following theorem, presented formally in Appendix A.3, shows a reduction to Single-Bounded from the NP-hard max-cover problem [12].

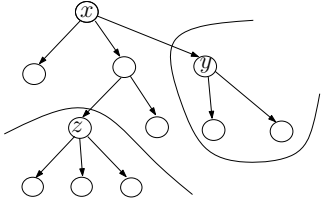

**Figure 3: Partition Example**

THEOREM 4.4. *Single-Bounded cannot be solved in polynomial time unless $P = NP$.*

Even though the general problem is intractable, it may be possible to find efficient solutions by leveraging specific characteristics of the input, and in particular of the DAG $G$. The following subsections examine this hypothesis for the two cases identified in the problem dimensions: downward-forests and upward-forests.

### 4.1.2 Single-Bounded: Downward-Forest

In the *downward-forest* case, $G$ is a forest of directed trees with edges from parents to children nodes (see also Section 2.)

We begin by showing that Single-Bounded on a downward-forest can be reduced to Single-Bounded on a downward-tree (a tree with edges from parents to children), by attaching a virtual root node that links all the trees in the forest. The following theorem is proved by showing that the optimal solution for the resulting tree gives a solution to the forest with wcase at most one more than optimal.

THEOREM 4.5 (DOWNWARD-FOREST ⇒ TREE). *Given a downward-forest $G_F$, there exists a downward-tree $G_T$, such that a solution $N_T$ to HumanGS on $G_T$ gives a node set $N'_T$ of $G_F$ such that $\text{wcase}(N'_T) \leq \text{wcase}(N) + 1$, for any node set $N$ of $G_F$ where $|N|, |N'_T| \leq k$.*

We can therefore focus on solving the Single-Bounded problem for a single downward-tree, instead of a downward-forest, ignoring the additive constant of $\leq 1$. We show that this problem is equivalent to the *partition problem* [19], which admits efficient solutions.

DEFINITION 4.6 (PARTITION PROBLEM). *Given an undirected tree, find $k$ edges such that their deletion minimizes the size of the largest connected component.*

To show the equivalence, we first define how a chosen set $N$ induces a partition of the tree into connected components.

DEFINITION 4.7 (PARTITION ON A NODE SET). *In a downward-tree, we recursively define the partitions on a node set $N$, denoted $P(N)$, as the following:*

- *If $x \in N$ and none of $x$'s descendants are in $N$, then the subtree under $x$ (including $x$) is a partition. (We call this partition the partition of $x$.)*
- *If $x \in N$ and some of $x$'s descendants are in $N$, then the subtree under $x$ (including $x$) excluding all of the partitions of $x$'s descendants is a partition. (We call this partition the partition of $x$.)*
- *Whatever is left after all the partitions are formed is a partition. (If $x$ is the root of the remainder of the tree, then the partition is called the partition of $x$.)*

Note that we have exactly $|N|$ or $|N|+1$ partitions. As an example, consider Figure 3. Here there are three partitions corresponding the node set $N = \{y, z\}$, i.e., the partition of $x$, $y$ and $z$.

LEMMA 4.8 (CANDIDATE SET PARTITION). *Given questions asked at a node set $N$, the candidate set $\text{cand}(N, u^*)$ for any $u^*$ corresponds to one of the partitions from $P(N)$.*

The proof of the lemma is given in Appendix A.5. The previous result essentially establishes the equivalence between the two problems, as our goal is to minimize $\text{wcase}(N)$, which is equivalent to the size of the largest partition that can be induced by $N$. Since the partition problem can be solved in PTIME, it follows directly that the same holds for Single-Bounded on a downward-tree. The following theorems formalize these observations.

THEOREM 4.9 (PARTITION PROBLEM EQUIVALENCE). *The problem of Single-Bounded on downward-trees is equivalent to the partition problem.*

Using a dynamic programming algorithm from [19] for the partition problem, we obtain the following result for Single-Bounded: (The proof of both these theorems is in the appendix.)

THEOREM 4.10 (SINGLE-BOUNDED). *There exists an algorithm with complexity $O(n \log n)$ that solves Single-Bounded on a downward-forest.*

### 4.1.3 Single-Bounded: Upward-Forest

Next, we assume that $G$ is an upward-forest. An upward-forest is a collection of upward-trees, where an upward-tree is a directed tree with directed edges from the children nodes to the parent.

We begin with a theorem that shows that it is sufficient to study upward-trees instead of upward-forests, by augmenting the forest with a new virtual root node such that there is an edge from each root node of each of the upward-trees to the new root node. The proof of the following result, sketched in Appendix A.9, is similar to that of Theorem 4.5.

THEOREM 4.11 (UPWARD-FOREST ⇒ TREE). *Given an upward-forest $G_F$, there exists a upward-tree $G_T$, such that a solution $N_T$ to HumanGS on $G_T$ gives a node set $N'_T$ of $G_F$ such that $\text{wcase}(N'_T) \leq \text{wcase}(N) + 2$, for any node set $N$ of $G_F$ where $|N'_T|, |N| \leq k$.*

Therefore, instead of upward-forests, we consider upward-trees. We visualize the tree with the root node at the top (with all edges going upward). This way, if we say a node $x$ has two children $y$ and $z$, we are disregarding the direction of the edges, which go from nodes $y$ and $z$ to $x$.

We use a dynamic programming algorithm, listed in Algorithm 2 in the appendix. Intuitively, the algorithm collects all possible worst-case contributions to the candidate set for the subtree under a certain node, and then combines the worst-cases when going from children to parent. For instance, the contribution to the overall candidate set when $x$'s subtree does not contain the target node is nothing but the sum of the contributions from the children of $x$, when neither of them have the target node in their subtree.

The following theorem establishes the correctness and complexity of the algorithm. The detailed proof of the theorem can be found in the Appendix A.10.

THEOREM 4.12 (UPWARD-FOREST). *There exists an algorithm that solves the Single-Bounded problem for $k$ questions in $O(m \cdot k^2 \cdot n^5)$ on an $m$-ary upward-forest with $n$ nodes.*

## 4.2 Single-Unlimited

In the Single-Unlimited problem, we do not have a strict budget on the number of questions that can be asked; instead, we want to find the smallest set of questions $N$ such that the target node is uniquely determined in the every case.

DEFINITION 4.13 (SINGLE-UNLIMITED). *(Unlimited Search for a single target node) Given that the target set $|U^*| = 1$, find the smallest set $N \subseteq V$ to ask questions such that $\text{wcase}(N) = 1$.*

Note that there are cases where the number of questions that are required to ensure that wcase($N$) = 1 can vary widely from $O(\log n)$ to $O(n)$. Additionally, we can repeat Single-Bounded with various values of $k$ to obtain a solution for Single-Unlimited for any graph. Both these results may be found in [4].

### 4.2.1 Single-Unlimited: Downward-Forest

On a downward-forest, Single-Unlimited can be solved in $O(n)$:

THEOREM 4.14 (DOWNWARD-FOREST). *On a downward-forest, we need to ask almost all nodes (except at most one) in order to solve Single-Unlimited.*

The proof may be found in Appendix B.1. The main insight in the proof is that if we leave a node unasked, then on getting a NO answer from all the children of the node and a YES answer from the parent of the node, we have wcase $> 1$ because we cannot distinguish between the node and its parent. However, in the general case of several trees in the forest, one of the roots need not be asked.

### 4.2.2 Single-Unlimited: Upward-Forest

The following theorem presents our result on Single-Unlimited on an upward-forest. The formal proof is in the extended technical report [4], but the main intuition is that we do not need to ask questions at any internal node with degree $> 1$, once all leaves have been asked. Note that the answer returned by every such internal node can be inferred from the answers of the children, bottom up. (If all the children return YES, then the node returns YES, else it returns NO.) Also, if we leave more than one leaf unasked, then there could be ambiguity regarding which of them is the target node.

THEOREM 4.15 (UPWARD-FOREST). *In order to solve Single-Unlimited on an upward-forest, we need to ask questions at all the leaves (except at most one), and all internal nodes with indegree 1, as in Algorithm 4 in the appendix.*

From the theorem we see that unlike the downward-forest case, we now do not need to ask questions at almost all nodes.

## 5. MULTIPLE TARGET NODES

In the Multi version of the problem, there exists a *target set* $U^* \subseteq V$, which denotes the unknown set of nodes we wish to discover by asking questions. Unlike the Single case, the size of the target set can be any $|U^*| \geq 1$ and is unknown. The only constraint we are given is that the target nodes satisfy the independence property, i.e., no two nodes in $U^*$ are related. We impose this constraint based on our motivating applications. (See Section 2.)

Computation of the candidate set for the Multi problem can be found in Section 3. To recap, a single question at $u$ lets us exclude from the candidate set either pset($u$) (if the answer is YES) or rset($u$) (if the answer is NO).

To incorporate the independence property in defining a worst-case candidate set, we define the function $ip$ on a set of nodes to return *true* if and only if the set of nodes satisfy the independence property. Given a set $N$ of nodes, we redefine the *Worst-Case Candidate Set* to be:

$$\text{wcase}(N) = \max_{U \subseteq V,\ ip(U)=\text{true}} |\text{cand}(N, U)|$$

Next, we study the Bounded version for the Multi case. The Unlimited version has a trivial solution (namely, $N = V$) and can be found in Appendix D.

### 5.1 Multi-Bounded

In this section we consider the bounded search problem for a target set of nodes, formally stated below.

DEFINITION 5.1 (MULTI-BOUNDED). *(Bounded Search for a target set.) Given a parameter $k$, find a set $N$ of nodes $N \subseteq V, |N| = k$ to ask questions such that wcase($N$) is minimized.*

We present algorithms and complexity results for Multi-Bounded, examining various properties of the structure of the graph $G$. Section 5.1.1 considers arbitrary DAGs, Section 5.1.2 considers downward and upward-forests.

### 5.1.1 Multi-Bounded: DAG

In this section, we establish the overall complexity of Multi-Bounded for an arbitrary DAG. We first show an NP-hard lower bound, and follow it with an upper bound of $\Sigma_2^P$. Bridging the gap between our lower and upper bounds is an open problem.

The following theorem establishes the NP-hardness of Multi-Bounded. As in Single-Bounded, we use the max-cover problem to prove NP-hardness, although the details of our reduction need to be modified for the Multi-Bounded problem. The proof of hardness can be found in Appendix C.1.

THEOREM 5.2 (LOWERBOUND). *The Multi-Bounded problem is NP-Hard in $n$ and $k$.*

The following theorem whose proof can be found in Appendix C.2 establishes the upperbound on the complexity of Multi-Bounded.

THEOREM 5.3 (UPPERBOUND). *The decision version of Multi-Bounded is in $\Sigma_2^P$.*

### 5.1.2 Multi-Bounded: Downward/Upward-Forest

Next we consider Multi-Bounded for forests of arbitrary trees. We can extend Theorems 4.5 & 4.11 to the case of Multi-Bounded, which enables us to focus our attention on trees instead of forests. Our first result shows that we do not need to consider upward and downward-trees separately. We then present the main result of this section, providing a PTIME dynamic programming algorithm that solves Multi-Bounded for downward-trees (and thereby downward and upward-forests).

THEOREM 5.4 (EQUIVALENCE). *Multi-Bounded on a downward-tree is equivalent to Multi-Bounded on an upward-tree.*

The proof can be found in Appendix C.3. Intuitively, if we reverse all the arrows, and we complement all the answers (replace YES with NO and NO with YES), we can transform Multi-Bounded on a downward-tree to that on an upward-tree and vice versa.

We use a dynamic programming algorithm for solving Multi-Bounded on a upward/downward-forest (Algorithm 5 in the appendix). This algorithm is similar to the one used to solve the Single-Bounded on an upward-forest.

THEOREM 5.5 (DP ALGORITHM). *There exists an algorithm that solves the Multi-Bounded problem for forests in $O(k^2 \cdot n^6)$.*

The details and proof of correctness of the theorem above can be found in the extended technical report [4].

## 6. EXPERIMENTAL STUDY

We conducted an experimental study of our HumanGS algorithms using a webpage categorization task on the real-world DMOZ concept hierarchy (http://dmoz.org). DMOZ is a human-curated internet directory based on a downward tree of categories. The goal is to assign websites of interest to nodes in this downward tree. In general, human judgement is needed for this assignment.

To expedite the process of assigning new web-pages to categories, we use the algorithms developed for HumanGS to select which questions to ask humans. A question at the node corresponding to category X would be of the form: does this new webpage fall under X? Note that questions for a given web-page can be asked concurrently to independent humans, and the answers can then be combined to determine the candidate set of categories.

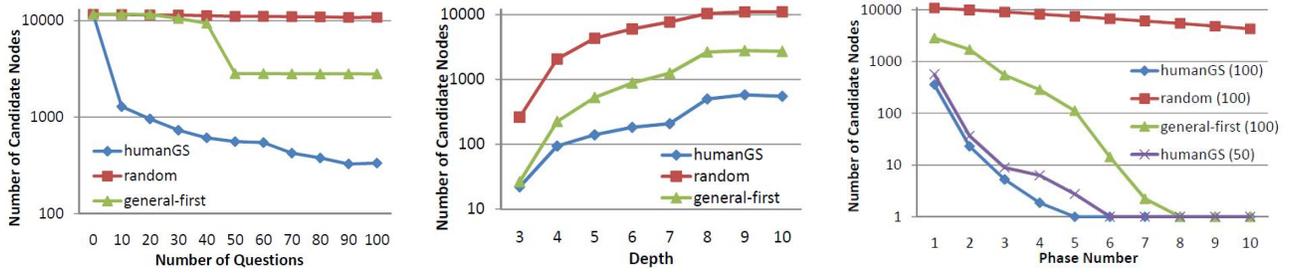

Figure 4: Experiments on (a) varying number of questions asked (b) varying the size of the tree (c) varying the number of phases

**Experimental Objectives.** Our experimental study has two important features that complement the analysis presented in previous sections. First, we study how our algorithms behave on average (rather than worst case) for a specific problem over real data. Second, since the algorithm for the Single-Unlimited problem for a downward tree is impractical, we study the performance when we use many iterations of the Single-Bounded algorithm. We now describe these features in more detail.

Recall that our algorithms are provably optimal in terms of minimizing the worst-case size of the candidate set. In our experiments, we would like to complement that theoretical analysis with measurements of the actual size of the candidate set after obtaining answers to the questions selected by the algorithm. By measuring the actual size, our goal is to examine whether the worst-case objective function also corresponds to good *average-case* performance.

One algorithm we could use for the webpage categorization task is Single-Unlimited. This algorithm selects questions offline to guarantee a wcase of size 1, but requires asking questions at almost every node, which is clearly impractical. Instead, since the average case may not correspond to the worst case, we would like to see how much the candidate set can be reduced by asking a bounded number of questions (i.e., using the algorithms for Single-Bounded). Using the answers to these questions, we may be able to precisely determine the target node. However, if the target node is not precisely determined, we may need to invoke the algorithm once again to issue another bounded set of questions to the crowdsourcing service. Thus, a practical method of using our algorithm is in *phases* where in each phase we invoke the Single-Bounded algorithm in order to select $k$ additional questions by taking the current candidate set as the input downward tree. The net effect is that each phase shrinks the candidate set further, until the target node is identified or the candidate set is small enough to assign to a human worker for the final solution. Note that this approach is a hybrid approach between a completely offline approach and a completely online approach (when we issue a single question at a time).

We present experiments to evaluate the average-case performance of our algorithms, focusing on the following questions:

- How does the candidate set shrink
  - as we vary the number of questions asked?
  - as we vary the size of the input downward tree?
  - when multiple phases are used?
- What is the relationship between the number of questions per phase and the total number of phases in order to solve the HumanGS task?

## 6.1 Methodology

**Task Specifics.** We evaluate our algorithms with a webpage categorization task on the science sub-tree of the DMOZ hierarchy (containing over 11,600 nodes). Each task is the placement of a web page into the hierarchy. We simulate each task by picking a node in the hierarchy where the web page would go. Then we simulate the crowd-sourcing service by answering the questions asked by our algorithms truthfully.

**Tested Algorithms.** We implemented our algorithm for Single-Bounded for a downward tree. We henceforth refer to this algorithm as *humanGS*. We compare *humanGS* against two baseline algorithms. The first algorithm, *random*, simply picks $k$ random nodes from the downward tree. The second algorithm, termed *general-first*, asks questions at the first $k$ nodes encountered in a breadth-first traversal starting from (but excluding) the root.

Recall that each algorithm is executed in phases, where each phase receives as input the graph formed from the candidate set computed using all previous phases.

**Metrics.** We measure the performance of an algorithm as the actual size of the candidate set after asking the questions selected by the algorithm. To ensure statistical robustness, we test each algorithm on a set of 100 random tasks, each generated by sampling the target node uniformly at random from the nodes of the input tree, and we report the average size of the candidate set over all tasks in the test set. For algorithm *random*, we perform an additional averaging step over 10 runs of the algorithm per test task, in order to mitigate the effects of random question sampling.

## 6.2 Results

**Effect of Number of Questions.** This experiment examines how the candidate set shrinks on increasing the number of asked questions in one phase. We use $k$ to denote the number of questions, and focus on the first phase where the input is the complete hierarchy.

Figure 4(a) shows the average number of candidate nodes as we vary $k$. Note that the y-axis is in log scale. As shown, our *humanGS* algorithm outperforms the baseline algorithms by an order of magnitude for all tested values of $k$. In particular, *humanGS* reduces the candidate set to around 1000 nodes with just 10 questions (a 10-fold decrease), whereas *random* and *general-first* flatten out well above 1000 nodes even after 100 questions.

**Effect of Tree Size.** The second experiment examines the performance of the three algorithms as we vary the size of the input tree. We varied the size of the downward tree by restricting the depth of the tree, i.e., any node at a greater depth is removed. The target node is sampled from the remaining nodes. We set $k = 50$ and focus again on a single phase.

Figure 4(b) depicts the average candidate size (again in log scale on the y-axis) against the depth of the tree. Algorithm *humanGS* continues to outperform its competitors, with an increasing margin as we approach the actual size of the tree.

**Phase-based Operation.** The final experiment evaluates the overall performance of the phase-based approach. Our goal is to examine how fast each algorithm identifies the single target node of the specific HumanGS task.

Figure 4(c) depicts the average candidate-set size for the three algorithms as we increase the number of phases, when $k = 100$ for each phase, once again with the y-axis on a log scale. The three plots corresponding to 100 questions per phase are denoted *humanGS (100)*, *random (100)* and *general-first (100)* in the figure. (We also plot the curve for *humanGS* when $k = 50$ and we discuss it in the next paragraph – depicted as *humanGS (50)* in the figure.) We observe that *humanGS* yields the fastest decrease among the three algorithms. For instance, *humanGS* is able to identify the target node after 5 phases on average. This compares favorably to the eight phases required by *general-first*, whereas *random* was not able to decrease the candidate set below 10000 nodes for the whole experiment. Additionally, the candidate-set size of *humanGS* is below 20 after only two phases.

Focusing on the two curves for *humanGS*, we observe a small degradation in performance from $k = 100$ to $k = 50$ for the same number of phases. However, $k = 50$ yields much better overall performance if we consider the total number of questions asked. For instance, with two phases and $k = 50$ (a total of 100 questions), *humanGS* performs an order of magnitude better than having a single phase with $k = 100$. Furthermore, *humanGS* requires 6 phases to discover the single target node with $k = 50$ (a total of 300 questions), compared to 5 phases for $k = 100$ (a total of 500 questions). These results indicate that increasing the number of phases is more beneficial compared to increasing the number of questions per phase. The trade-off of course is in latency, since the extra phases mean additional round-trips to the crowd-sourcing service, and reduced parallelism. Examining this trade-off in more detail is an interesting direction for future work.

## 7. RELATED WORK

There have been several studies of the social aspects of crowd-sourcing technologies and on how to design social games (e.g., captchas and GWAP — Games With A Purpose) in order to accomplish tasks [2, 11]. Previous works have also examined the use of crowd-sourcing technologies to accomplish specific tasks, such as natural language annotations [16], video and image annotations [10, 5], and search relevance [14]. However, none of these studies consider the problem of minimizing the set of questions to ask users; additionally, the tasks studied are fairly simple and unstructured, unlike HumanGS which (a) utilizes the inherent structure of a DAG to ask questions, and (b) requires asking multiple questions in order to be "solved" correctly.

The field of active learning [6] also studies the problem of requesting human input. The goal is to request input from experts with the maximal "information content", similar in spirit to our problem. However, this input is only used to generate training data for machine learning tasks (especially when the current data is insufficient or not informative). Additionally, most work in active learning does not ask questions to humans in parallel, and thus does not leverage the inherent parallelism in crowd-sourcing systems.

There has been some work on the utilization of human input in data integration and exploration [18, 22, 20]. Similar to HumanGS, the aim is to minimize the number of posed questions in order to minimize the total cost. However, the models and applications studied are very different from ours, and not as generally applicable. In addition, the tasks are such that the human inputs "assist" the computer program. In our case, the tasks are fairly complex, as a result of which they require primarily human input to solve them.

Our problem is similar to that of decision trees [7]. In decision trees, we wish to classify a target item as belonging to one of many classes (in our case, classify a target item as being one of the nodes in the DAG). However, unlike decision trees, we do not have a training set (or statistics of various classes) and there are no attributes on which a classifier can be built. The only questions that we can ask are those that involve reachability, and the optimization issues that arise are very different in our case.

## 8. CONCLUSIONS AND FUTURE WORK

In this work, we presented HumanGS, a general problem that arises in many problems of human computation. We explored the problem space via three orthogonal axes: Single / Multi, Bounded / Unlimited and DAG / Downward-Forest / Upward-Forest, and developed algorithms for all combinations. Our algorithms generate the optimal set of questions that can be asked to humans. These questions do not rely on the answers received for prior questions, and can be issued in parallel in a crowd-sourcing system.

Since optimization of human computation is an unexplored area, there are many interesting avenues of future work. (See [3] for additional open problems.) Within HumanGS itself, one interesting direction is to assume that we are given a probability distribution on the selection of target nodes, which we can use to pick questions in order to minimize the expected size of the candidate set. Additionally, some questions asked to humans may be answered incorrectly, and in such a case we would like to perform error-resilient HumanGS. Last, but not the least, we would like to explore other problems in the human computation space that require optimization by way of picking the optimal questions to ask humans.

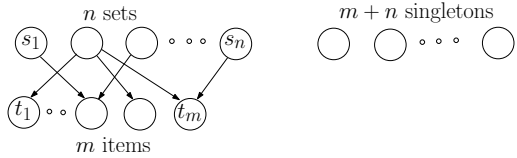

**Figure 5: Hardness Proof for Single-Bounded**

# APPENDIX
## Outline

We cover the details of Single-Bounded in Appendix A, Single-Unlimited in B, Multi-Bounded in C and Multi-Unlimited in D.

In the appendix, we provide

- the complete analysis of Multi-Unlimited, and
- the analysis of balanced trees for Single and Multi-Bounded

as well as:

- proofs for some theorems and
- pseudocode for the algorithms

from the main body of the paper.

## A. Single-Bounded

### A.1 Proof of Theorem 4.2

PROOF. In time $O(n^2)$, for all pairs of nodes $a$ and $b$ in $G$, we can compute and maintain whether or not $a \in \mathsf{rset}(b)$. Let $u^*$ be the target node in the graph. In time $O(|N|)$, we can compute the answers to each of the questions asked at the nodes in $N$. Subsequently, we can compute $\mathsf{cand}(N, u^*)$ in $O(n \cdot |N|)$ by checking the following for every node $u$: if there is a node in $u' \in \mathsf{rset}(u), u \neq u'$ such that $u' \in N$ and $u'$ returned YES, or if there is a node $u \in \mathsf{pset}(u)$, such that $u \in N$ and $u'$ returned NO, then $u \notin \mathsf{cand}(N, u^*)$, else $u \in \mathsf{cand}(N, u^*)$. We can compute wcase by repeating the above procedure for every possible $u^*$, taking a total of $O(n^2 \cdot |N|)$ time. □

### A.2 Proof of Lemma 4.3

PROOF. We find wcase for each choice of $\binom{n}{k}$ nodes at which questions are asked. The choice for which the worst-case candidate set is the smallest is the optimal solution. □

### A.3 Proof of Theorem 4.4

PROOF. We prove that the decision version of the problem is NP-complete, stated as follows: Given a budget $k$ and a positive integer $m$, is there a node-set $N$ to ask questions such that $\mathsf{wcase}(N) \leq m$? We refer to this problem as Single-Bounded-DECISION.

We use a reduction from the NP-complete *max-cover problem* [12]. In this problem, the objective is to pick a certain number of sets such that they cover as many items as possible. (A set containing a given item is said to cover that item.) Let there be $m$ items and $n$ sets in the max cover problem. We need to select $k$ sets such that the maximum number of items are covered.

We reduce the max-cover problem to Single-Bounded-DECISION with the following directed acyclic graph. (An instance of the graph is shown in Figure 5.) Consider nodes arranged in two layers. In the first layer, we have one node corresponding to each set $s_i$ in the max cover problem. In the second layer, we have a node corresponding to each item $t_i$ in the max-cover problem. There is a directed edge from the node corresponding to set $s_i$ to the node corresponding to the item $t_j$ iff item $t_j$ is present in set $s_i$. In addition, we include $n+m$ singleton nodes (nodes with no incoming or outgoing edges) in the DAG. Subsequently, we call Single-Bounded on this DAG with a budget of $k$ questions. The worst-case candidate set corresponds to each of the questions in $N$ receiving NO answers. To see this, note that if we get a YES answer to a question asked at any of the nodes corresponding to sets, the candidate set is $\leq m$. Receiving a YES answer to any question asked at the nodes corresponding to items or singletons will give a candidate set of 1. However, if we get all NO answers, the number of nodes remaining in the worst-case candidate set is at least $m+n$, even if all the nodes corresponding to items as well as $k$ from the nodes corresponding to sets and singletons are eliminated from the candidate set. Additionally, the solution of the Single-Bounded problem, i.e., $N$, will only contain nodes corresponding to sets, because those nodes exclude the maximum number of nodes from the candidate set, in the worst case. Thus, Single-Bounded picks nodes corresponding to sets such that the maximum number of nodes corresponding to items are covered. Therefore, the solution to Single-Bounded corresponds to a max-cover. Conversely, every solution of the max-cover problem can be written as a solution for Single-Bounded.

In addition, given that we can compute $\mathsf{wcase}(N)$ in PTIME (see Theorem 4.2), a solution for Single-Bounded-DECISION can be verified in PTIME. Thus, Single-Bounded-DECISION on DAGs is in NP and is NP-Complete. □

### A.4 Proof of Theorem 4.5

PROOF. We augment the downward-forest with a single root node such that there exists an edge from the new root node to the root of each of the trees in the directed forest. Let the original forest be $F$ and the new augmented tree (a downward-tree) be $T$.

Let $N_T$ be an optimal selection of $k$ nodes from $T$ at which questions are asked. We first convert $N_T$ into $N'_T$ such that there is no question asked at the root. (We delete the root from $N_T$, if present.) Since a question at the root has to return a YES answer, the worst-case candidate set at $T$, denoted by $\mathsf{wcase}_T$, will be unchanged. We therefore have $\mathsf{wcase}_T(N'_T) = \mathsf{wcase}_T(N_T)$.

Let $N_F$ be the optimal solution on $F$. If we select set $N_F$ as the questions to be asked at $T$, we get $\mathsf{wcase}_T(N_T) \leq \mathsf{wcase}_T(N_F)$, since $N_T$ gives the optimal worst-case candidate set for $T$. Also, since there is an addition of at most one node to the worst-case candidate set when $N_F$ is used on $T$ instead of $F$, $\mathsf{wcase}_T(N_F) \leq \mathsf{wcase}_F(N_F) + 1$, where $\mathsf{wcase}_F$ is the worst case candidate set when the questions are asked at the nodes in $F$. Now, apply $N'_T$ on $F$. We then have $\mathsf{wcase}_F(N'_T) \leq \mathsf{wcase}_T(N'_T) = \mathsf{wcase}_T(N_T) \leq \mathsf{wcase}_F(N_F)+1$. Thus, choosing the questions to minimize worst-case candidate set in the tree gives the optimal worst-case candidate set for the forest plus at most one more node. We ignore the additive factor of 1 in our calculations. □

### A.5 Proof of Lemma 4.8

PROOF. If $u^* \in$ partition $p$ of $x$, then we prove that $\mathsf{cand}(N, u^*) = p$. First, we consider the case when $x \in N$. In this case, notice that the question at $x$ returns a YES answer. Using Theorem 3.3, the candidate set can be restricted to the subtree under $x$. The only nodes $\in N$ at which the answer is YES are those that are ancestors of $x$, but they do not change the candidate set. Those nodes $\in N$ that are descendants of $x$ or unrelated to $x$ all return NO, since there is no path from those nodes to the target node $u^*$. Since all those nodes return NO, the subtrees under those nodes can be removed from the candidate set. Thus, the candidate set is precisely $p$.

If $x \notin N$, then $x$ is the root of the directed tree. Here, once again, if we asked a question at $x$, we would obtain a YES answer. Those nodes $\in N$ that are descendants of $x$ all return NO, since there is no path from those nodes to the target node $u^*$. Since all those nodes return NO, the subtrees under those nodes can be removed from the candidate set. Thus, the candidate set is once again $p$. □

## A.6 Proof of Theorem 4.9

PROOF. As seen in Lemma 4.8, the candidate set on asking a question at any node corresponds to one of the partitions induced by asking questions. Thus, in order to minimize wcase, it is sufficient to solve the partition problem on the downward-tree. The size of the largest partition is the size of the worst possible candidate set. Conversely, every instance of the partition problem can be cast as an instance of the Single-Bounded problem for a downward-tree. Thus, the two problems are equivalent. □

## A.7 Proof of Theorem 4.10

PROOF. The algorithm solves the equivalent partition problem on the downward-tree formed from the downward-forest augmented with a root node. Reference [19] gives a dynamic programming algorithm that finds the minimal number of edges that need to be cut in order to achieve a certain partition size in $O(n)$. We run binary search over partition sizes in order to find the smallest partition size for which the minimal number of edges to be cut is $\leq k$. □

## A.8 Balanced Downward-Trees

**Algorithm 1:** Single-Bounded Balanced Downward-Tree
**Data**: $\mathcal{G} = m$-ary balanced downward-tree with depth $d$, $k$
**Result**: optimal set of $k$ nodes to ask questions to
$l := \lfloor \log_m k \rfloor$;
$N := $ all nodes at depth $l$;
**if** $k \neq m^l$ **then**
  **for** all nodes $x$ at depth $l$ **do**
    $N := N \cup $ first $\lfloor (k - m^l)/m^l \rfloor$ children of $x$;
return $N$;

For a single balanced tree, the algorithm is listed in Algorithm 1 when $k < m^{d/2}$. Intuitively, the algorithm first picks all nodes at depth $l$, where $l$ is the maximum such depth that can be "blanketed" with questions. Once all nodes at this depth is added to $N$, the algorithm then proceeds to sub-divide each of the subtrees at depth $l$ (i.e., those whose roots are at depth $l$) by asking further questions, which are equally distributed among the subtrees at depth $l$. The detailed discussion and proof of optimality of the following theorem can be found in the extended technical report [4].

THEOREM A.1 (BALANCED DOWNWARD-TREE). *For a balanced downward $m$-ary tree of depth $d$, Single-Bounded can be solved optimally using Algorithm 1 in $O(1)$ if $k < m^{d/2}$.*

## A.9 Proof of Theorem 4.11

PROOF. We augment the upward-forest with a single root node such that there exists an edge to the new root node from the root of each of the trees in the directed forest. Let the original forest be $F$ and the new augmented tree (a upward-tree) be $T$. The proof is similar to that of Theorem 4.5, except that instead of being equal, $\text{wcase}_T(N'_T) \leq \text{wcase}_T(N_T) + 1$, due to which we have $\text{wcase}_F(N'_T) \leq \text{wcase}_T(N'_T) \leq \text{wcase}_F(N_F) + 2$. Additional details may be found in the extended technical report [4]. □

## A.10 Upward-Forests

First, we define the subtree of $x$ to be $x$ along with all nodes below $x$, i.e., all $x'$ such that $x \in rset(x')$.

We use a dynamic programming algorithm[2], listed in Algorithm 2, but the steps are explained next. The algorithm builds an array

---
[2]While the algorithm is only described for 2-ary trees, it can be easily generalized to handle $m$-ary trees.

**Algorithm 2:** Single-Bounded Upward-Forest
**Data**: $\mathcal{G} = $ upward-tree, $k = $ budget
**Result**: optimal set of $k$ nodes to ask questions to
**for** all nodes $x$ in $\mathcal{G}$, bottom-up **do**
  $T_x := \emptyset$;
  $T_x[0] := \{((0,0), (size(x), 0), 0, size(x))\}$;
  **if** $x$ has 1 or 2 children **then**
    $y := $ left sub-child of $x$;
    $z := $ right sub-child of $x$;
    **for** $i : 0 \ldots k$ **do**
      **for** all $k_1, k_2 : k_1 + k_2 = i$ **do**
        **for** all $((*,*), (p_1, p_2), n, r) \in T_y[k_1]$ and all $((*,*), (p'_1, p'_2), n', r') \in T_z[k_2]$ **do**
          $p_b := \max\{p_1, p_2 + n', p'_1, p'_2 + n\}$;
          $n_a := n + n'$;
          $r_a := p_a := 1$;
          **if** $k_1 == 0 \land k_2 \neq 0$ **then**
            $p_a := \max(p'_1 + 1, r' + 1)$;
            $p_b := \max\{n', p'_2\} + size(y)$;
            $r_a := r' + 1$;
          **if** $k_2 == 0 \land k_1 \neq 0$ **then**
            $p_a := \max(p_1 + 1, r + 1)$;
            $p_b := \max\{n, p'_2\} + size(z)$;
            $r_a := r + 1$;
          **if** $x$ has one child $y$ **then**
            $T_x[i+1] := T_x[i+1] \cup \{((i,0), (1, \max\{p_1, p_2\}), n_a, 1)\}$;
            **if** $i \neq 0$ **then**
              $T_x[i] := T_x[i] \cup \{((i, 0), (p_a, p_2), n_a, r_a)\}$;
          **else**
            $T_x[i+1] := T_x[i+1] \cup \{((k_1, k_2), (1, p_b), n_a, 1)\}$;
            **if** $i \neq 0$ **then**
              $T_x[i] := T_x[i] \cup \{((k_1, k_2), (p_a, p_b), n_a, r_a)\}$;
      compress $T_x[i]$;
  **else**
    **for** $i : 1 \ldots k$ **do**
      $T_x[i] := \{((k, 0), (1, 0), 0, 1)\}$;
$r := $ root of tree;
$t := $ tuple in $T_r[k]$ that has smallest $p_1$;
trace the origin of $t$ until the leaves of the tree;
output the questions;

$T_x[i]$, for each node $x \in V$, and $i \in \{0, \ldots, k\}$, bottom-up. Intuitively, in $T_x[i]$, we maintain a set of options for asking questions at the children $y$ and $z$ of $x$, by recording the worst-case contributions to the candidate set for various locations of the target node for each option, as described below:

Formally, we define array $T_x[i]$ to contain a set of 4-tuples of the following form $(K, P, N, R)$, where $P$ and $K$ are pairs of values while $N$, $R$ are singleton values. Consider a 4-tuple $((k_1, k_2), (p_1, p_2), n, r)$: This 4-tuple indicates that when $i$ questions are allocated to node $x$, there is a configuration of asking questions in $x$'s subtree, along with $k_1$ questions in $y$'s subtree, and $k_2$ questions in $z$'s subtree, such that no matter where the target node is, the worst possible contribution to the overall worst-case candidate set coming from $x$'s subtree corresponds to one of $p_1, p_2, n, r$. That is, if we were to ask questions at a node set $N$ in $x$'s subtree, the worst contribution from $x$'s subtree to the overall candidate set would be one of these numbers. Value $p_1$ corresponds to the worst contribution such that (a) it contains the root $x$ and (b) the target node is present

**Algorithm 3:** Single-Bounded Balanced Upward-Tree

**Data**: $\mathcal{G} = m$-ary balanced upward-tree with depth $d$, $k$
**Result**: optimal set of $k$ nodes to ask questions to
$\alpha := \lfloor \log_m k \rfloor$;
**if** $k \neq m^\alpha$ **then**
  $\alpha := \alpha + 1$;
$N := \emptyset$;
**while** $|N| < k$ **do**
  pick a new node $n$ at depth $\alpha$;
  pick a leaf $l$ in $n$'s subtree;
  $N := N \cup \{l\}$;
return $N$;

---

in $x$'s subtree, $p_2$ corresponds to the worst contribution such that (a) it does not contain $x$ and (b) the target node is present in $x$'s subtree, $n$ corresponds to the contribution when the target node is neither present in $x$'s subtree nor is in $\mathsf{rset}(x)$, (There is only one number as against two for P, since the contribution to the candidate set will never contain the root.) while $r$ corresponds to the contribution when the target node is not present in $x$'s subtree but is reachable from $x$. (There is only one number since the contribution to the candidate set will always contain the root.)

The array $T_x[i]$ is thus a collection of worst-case tuples. In the worst case, we might need to maintain upto $O(n^4)$ tuples (corresponding to all combinations of the last 3 entries in the tuple). However, we can also use the following rule to discard tuples: If all values in the last three entries in the first tuple are greater than or equal to the corresponding values in the second tuple, we can discard the first tuple in favor of the second because the second allocation is better overall than the first. This approach corresponds to maintaining the minimum-skyline of these tuples. However, this procedure is not necessary for correctness.

The detailed proof, including consideration of all cases can be found in [4].

## A.11 Balanced Upward-Trees

We now present an efficient algorithm for an balanced upward-tree. The following theorem formalizes our results.

THEOREM A.2 (BALANCED UPWARD-TREE). *For a balanced upward $m$-ary tree of large depth $d$, Algorithm 3 finds the optimal solution for the Single-Bounded problem in $O(1)$.*

The proof of the theorem can be found in the extended technical report. Essentially, the theorem tells us that in order to pick questions in balanced upward-trees, we should pick "leaf" questions. Intuitively, asking a question at a leaf node gives us the maximum amount of information about the location of the target node. (In particular it lets us eliminate the entire path to the root from the worst-case candidate set.) However, we must pick leaf nodes that are as "spread out" as possible, and share as little of their path to the root as possible. For more details, refer Algorithm 3.

## B. Single-Unlimited
### B.1 Proof of Theorem 4.14

PROOF. Consider all leaf nodes across all trees in the forest. Note that each leaf node has a single parent. Firstly, it is easy to see that we need to ask all leaves. If not, consider a leaf node $a$. Let its parent be $b$. If we do not ask $a$, but $b$ returns YES, (but no other child of $b$ returns YES) then it is not clear if the target node is $a$ or $b$. Therefore, we need to ask $a$. The same argument can be used for all leaves. Now assume all questions at leaves return a NO answer (effectively, leaves can be removed from the tree.) The argument can be repeated for each parent of the leaves as well.

---

**Algorithm 4:** Single-Unlimited Upward-Tree

**Data**: $\mathcal{G} = m$-ary upward-tree, $k$
**Result**: optimal set of $k$ nodes to ask questions to
**if** $\exists$ a leaf $f$ such that $f$'s parent has degree $\neq 2$ **then**
  $N = $ all leaves except $f$;
**for** all internal nodes $x$ **do**
  **if** $x$ has indegree 1 **then**
    $N = N \cup \{x\}$;
return $N$;

---

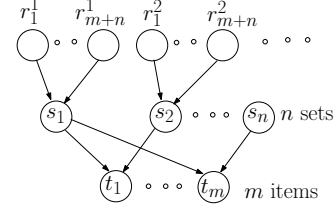

Figure 6: Hardness Proof for Multi-Bounded

We then arrive at the roots of each of the trees in the forest and their children. We cannot leave two of the roots unasked because we cannot distinguish between them without asking a question at both of them. However, we can avoid asking a question at one of the roots because if we get a NO response from all other trees as well as children of that node, then that node has to be the target node. □

## C. Multi-Bounded
### C.1 Proof of Theorem 5.2

PROOF. We give a reduction from the NP-hard max-cover problem: Given $m$ items, $n$ sets, and an integer $k$, the goal is to choose $k$ sets that cover the most number of items.

We only describe the construction here. The proof that the construction is a reduction can be found in the extended technical report [4]. Given an instance of the max-cover problem, we construct an instance of Multi-Bounded with the following DAG containing three layers of nodes, as depicted in Figure 6. The second and third layers of nodes are identical to the first and second layers in the proof for Theorem 4.4: The second layer has one node corresponding to each set and the third layer has a node corresponding to each item. There is an edge from node $s_i$ to the node $t_j$ iff item $t_j \in s_i$ in the max-cover instance. For each node $s_i$ in the second layer, we add $m + n$ unique parents in the first layer, $r_1^i, r_2^i, \ldots, r_{m+n}^i$, with an edge to $s_i$. We want to solve Multi-Bounded on this constructed DAG for $k$ questions. Note that to solve the Multi-Bounded problem, we will always pick nodes corresponding to sets, because they let us eliminate the maximum number of nodes corresponding to items, in the worst case. Thus the nodes corresponding to sets that are picked in Multi-Bounded precisely correspond to the sets that are picked in the max cover problem. □

### C.2 Proof of Theorem 5.3

PROOF. Given an instance of the decision version of the Multi-Bounded problem, we can express it as an instance of $\Sigma_2^P$ in the following way:

$$\exists y_1 \forall y_2 [L(y_2) \lor (R(y_1, y_2) < X)],$$

where $y_1$ corresponds to a set of nodes at which questions are asked, $y_2$ corresponds to all possible instances of the target set. $L(y_2)$ checks in PTIME whether $y_2$ contains two nodes with a path from one to the other: If so, it returns YES. $R(y_1, y_2)$ evaluates the candidate set given $y_1$ and $y_2$. □

**Algorithm 5:** Multi-Bounded Downward/Upward-Forest

**Data**: $\mathcal{G} =$ upward-tree, $k =$ budget
**Result**: optimal set of $k$ nodes to ask questions to
**for** all nodes $x$ in $\mathcal{G}$, bottom-up **do**
    $T_x := \emptyset$;
    $T_x[0] := \{((0,0), size(x), 0, size(x))\}$;
    **if** $x$ has 1 or 2 children **then**
        $y :=$ left sub-child of $x$;
        $z :=$ right sub-child of $x$;
        **for** $i : 0 \ldots k$ **do**
            **for** all $k_1, k_2 : k_1 + k_2 = i$ **do**
                **for** all $((*,*), p_1, p_2, n)$ in $T_y[k_1]$ and all $((*,*), p'_1, p'_2, n')$ in $T_z[k_2]$ **do**
                      $p_a := \max\{p_1 + n', p'_1 + n, p_1 + p'_1, n + n'\} + 1$;
                      $p_b := \max\{p_2 + p'_2, p_2 + p'_1, p'_2 + p_1, p_2 + n', p'_2 + n\}$;
                      $n_a := n + n' + 1$;
                      **if** $k_1 == 0 \land k_2 \neq 0$ **then**
                          $p_b := p'_2 + n$;
                      **if** $k_2 == 0 \land k_1 \neq 0$ **then**
                          $p_b := p_2 + n'$;
                      $T_x[i+1] := T_x[i+1] \cup \{((k_1, k_2), p_a, p_b, 0)\}$;
                      **if** $i \neq 0$ **then**
                          $T_x[i] := T_x[i] \cup \{((k_1, k_2), p_a, p_b, n_a)\}$;
            compress $T_x[i]$;
    **else**
        **for** $i : 1 \ldots k$ **do**
            $T_x[i] := \{((k, 0), 1, 0, 0)\}$;
$r :=$ root of tree;
$t :=$ tuple in $T_r[k]$ that has smallest $p_1$;
trace the origin of $t$ until the leaves of the tree;
output the questions;

## C.3 Proof of Theorem 5.4

PROOF. Consider the answers to a set of questions asked at nodes in a downward-tree. If we were to complement each answer (YES → NO and NO → YES), we would obtain the answers to the questions asked at nodes in the same tree if the direction of each edge was reversed (i.e., the upward-tree) for the same target set. (Recall from Theorem 3.3 that each question either rules out the ancestors or the descendants from the candidate set.) Thus, we can solve Multi-Bounded on a downward-tree by reversing all edges and solving it on the upward-tree. The converse is also true. Hence the problems are equivalent. □

## C.4 Balanced Trees

We now consider solving Multi-Bounded for the special case of balanced trees. (Recall we don't need to consider upward and downward-trees separately.) The following theorem shows that for small $k$ (compared to the number of nodes in the tree), there is a constant-time solution for a balanced $m$-ary tree.

THEOREM C.1 (BALANCED TREES). *Multi-Bounded on a balanced $m$-ary tree of height $d$ is solvable in $O(1)$ when $k \leq m^{\frac{d}{2}}$.*

The proof may be found in the extended technical report [4], while the algorithm is listed in Algorithm 6. Informally, the approach is to pick nodes at a depth such that the path to the root is equal to the size of the sub-tree underneath (such that a YES answer or a NO answer would eliminate roughly the same number of nodes from candidate set), in order to minimize wcase.

**Algorithm 6:** Multi-Bounded Balanced Trees

**Data**: $\mathcal{G} = m$-ary balanced tree with depth $d$, $k$
**Result**: optimal set of $k$ nodes to ask questions to
$l := \lfloor \log_m k \rfloor$;
$N := \emptyset$;
**while** $|N| < k$ **do**
    pick a new node $n$ at depth $l$;
    pick a node $x$ at level $\alpha$ in subtree under $n$;
    $N := N \cup \{x\}$;
return $N$;

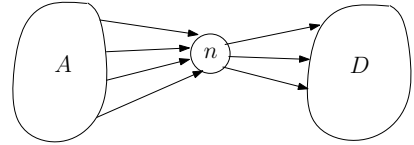

**Figure 7: Triviality of Multi-Unlimited**

## C.5 Balanced Forest

We can generalize the results for balanced trees to obtain a better solution for *balanced forests*, i.e., a collection of balanced trees. The generalization to forests employs a dynamic programming algorithm operating over the balanced trees constituting the forest.

THEOREM C.2 (BALANCED FOREST). *Multi-Bounded for a balanced forest can be solved in $O(k^2 r)$ where $r$ is the number of trees.*

The proof of this theorem can be found in the extended technical report [4]. The algorithm iteratively (for each $i$) computes the best allocation of questions to each of the balanced trees, by using the optimal allocation of $k_1$ questions to the first $i$ balanced trees, and $k_2$ questions to the $(i+1)$th tree, and combining the worst-case candidate sets (The candidate sets for each balanced tree are independent and can be combined.) to give one candidate allocation of $k_1 + k_2$ questions to the first $i+1$ trees. We compute this value for all possible $k_1$ and $k_2$ such that $k_1 + k_2 \leq k$, and all $i$, recursively, and return the best possible allocation when $i = r$.

## D. Multi-Unlimited

In this section, we address the problem of Multi-Unlimited.

DEFINITION D.1 (MULTI-UNLIMITED). *(Unlimited Search in a DAG for a target set) Find the smallest set of nodes $N \subseteq V$ to ask questions such that $\forall U^* \subseteq V$ satisfying $ip(U^*) = 1$, we have $|\text{cand}(N, U^*)| = |U^*|$.*

The following theorem shows an interesting result that the Multi-Unlimited problem is "trivialized" by the fact that questions need to be asked at all nodes to ensure that no extraneous nodes remain in the candidate set.

THEOREM D.2 (TRIVIALITY). *The optimal solution to an instance of Multi-Unlimited is $N = V$, i.e., we need to ask a question at every node in the graph.*

PROOF. Consider Figure 7, abstractly representing a connected component of the input graph, focusing on any node $n$. We prove that we need to ask a question at $n$ in order to ascertain if $n \in U^*$. Suppose we don't ask a question at $n$. Let the questions asked at all of the ancestors of $n$, i.e., $A$, return YES, while questions at all of the descendants of $n$, i.e., $D$, return NO. In this case, it is not clear if $n$ is in $U^*$ or not. It is possible that $n$ is in $U^*$, in which case none of the nodes in $A$ form part of $U^*$. Otherwise, if $n \notin U^*$, then there may be many nodes from $A$ which are part of $U^*$. □